\documentclass[aps,prl,showpacs,nobibnotes,twocolumn,
nobalancelastpage,superscriptaddress]{revtex4}
\usepackage{graphicx}
\usepackage{amsmath}
\usepackage{latexsym}
\usepackage{dcolumn}
\usepackage{bm}
\input{epsf}
\newcommand{\be}{\begin{equation}}
\newcommand{\ee}{\end{equation}}
\newcommand{\ba}{\begin{eqnarray}}
\newcommand{\ea}{\end{eqnarray}}
\newcommand{\bi}{\begin{itemize}}
\newcommand{\ei}{\end{itemize}}
\newcommand{\ga}{\gtrsim}
\newcommand{\bfi}{\begin{figure}
\epsfxsize=9cm
\epsffile}
\newcommand{\efi}{\end{figure}}
\newcommand{\la}{\lesssim}
\newcommand{\mpch}{h^{-1} {\rm Mpc}}
\newcommand{\gpch}{h^{-1} {\rm Gpc}}

\begin{document}
\title{The exact analytical solution of the linear structure growth rate in
  $\Lambda$CDM cosmology and its cosmological applications}
\author{Pengjie Zhang}
\email{pjzhang@shao.ac.cn}
\affiliation{Key laboratory for research in galaxies and cosmology, Shanghai Astronomical Observatory, Chinese Academy of
  Science, 80 Nandan Road, Shanghai, China, 200030}
\begin{abstract}
We derive the exact analytical solution of the linear structure growth rate in
$\Lambda$CDM cosmology with flat or curved geometry, under the Newtonian gauge. Unlike the well known
solution under the Newtonian limit (Heath 1977, \cite{Heath77}),  our solution takes all
general relativistic corrections into account and is hence valid at both the
sub- and super-horizon  scales. With this exact solution, we evaluate
cosmological impacts induced by these relativistic corrections.  (1) General
relativistic corrections alter the density growth from $z=100$ to $z=0$ by $10\%$ at
$k=0.01h/$Mpc and the impact becomes stronger toward larger scales. We
caution the readers that the overdensity is not gauge invariant and the above
statement is restrained to the Newtonian gauge.  (2) Relativistic corrections  introduce a  $k^{-2}$ scale dependence
in  the density fluctuation. It mimics a primordial non-Gaussianity of the local type
with $f^{\rm local}_{\rm  NL}\sim 1$. This  systematical error may become
non-negligible for  future all sky deep galaxy surveys. (3) Cosmological
simulations with box size greater 
than 1Gpc are also affected by these relativistic corrections. We provide a
post-processing recipe  to correct for these effects. (4) These
relativistic corrections affect the redshift distortion. However, at
redshifts and scales relevant to redshift distortion measurements, such effect
is negligible. 
\end{abstract}
\pacs{98.65.Dx; 04.25.Nx}
\maketitle

\section{Introduction}
A well known and widely used solution of the linear density growth in
$\Lambda$CDM cosmology, first derived
by Heath 1977 \cite{Heath77},  is
\be
\label{eqn:Heath77}
\delta_m\propto D_{m,N}\propto H\int_0^a \frac{da}{H^3a^3}\ .
\ee
Here, $\delta_m$ is the matter overdensity. $D_{m,N}$ is the linear density
growth rate in the Newtonian limit. Throughout the paper, the subscript
``N'' denotes the corresponding property in the Newtonian limit. $H=H(a)$ is the Hubble parameter at redshift $z=1/a-1$ and $a$ is the
scale factor. This solution is derived under the limit of
negligible radiation, negligible baryon pressure and  identical initial
conditions for fluctuations in baryons and dark matter. These conditions are
adopted throughout the paper. 

This solution is
valid for arbitrary cosmological constant 
$\Lambda$. It is also valid for arbitrary curvature $K$, at scales much
smaller than the curvature radius
$r_K=1/\sqrt{K}=1/H_0\sqrt{|\Omega_K|}$. Here 
$\Omega_K$ is the dimensionless curvature density and $H_0\equiv H(a=1)$ is
the present day Hubble constant.  Throughout the paper, we
set the speed of light $c=1$, so $1/H_0=3\gpch$. 

However, this solution is derived by neglecting relativistic corrections
to the Poisson equation and to the continuity equation, so it is valid only in the
Newtonian limit and hence at sub-horizon scale, where the wavevector $k$
satisfies
\be
\label{eqn:horizon}
k\gg aH=\frac{a\sqrt{\Omega_0a^{-3}+\Omega_{\Lambda}+\Omega_Ka^{-2}}}{3\times 10^3 \mpch}
\ .
\ee
Here, $\Omega_0$ and $\Omega_\Lambda$  are the dimensionless matter and 
$\Lambda$ today, respectively.  

Modern surveys are pushing the observational boundary to  larger scale and
higher redshift. This enables several important cosmological
applications. (1) Galaxy clustering at scales $k\la 0.01 h/$Mpc is a sensitive 
probe of primordial non-Gaussianity (PMG). It relies on the asymptotic
behavior of galaxy clustering at $k\rightarrow 0$ to probe PMG
(e.g. \cite{Dalal08}). (2) The large scale structure (LSS) growth history, often
inferred  by  combining low redshift and high redshift measurements, contains key
information to test general relativity and probe dark energy. Such measurement
is usually done at $k\ga 0.01 h/$Mpc  scale. However, since the horizon scale 
decreases with increasing redshift, even a mode deep in the  sub-horizon
regime today can be close to the horizon scale at earlier epoch. 

For these cosmological applications, whether
relativistic corrections can be safely neglected and whether
Eq. \ref{eqn:Heath77} is sufficiently accurate are  becoming issues of practical
importance and have evoked discussions (e.g. \cite{Dent09}).  Existing 
packages such as CMBFAST \cite{CMBFAST} and CAMB \cite{CAMB} numerically solve 
the fully general relativistic  linear perturbation equations. So they can be
reliably resorted to calculate the structure growth  at all linear scales.

 However,  it is still useful to understand the role of
these relativistic corrections  through  analytical
approach.  This motivates us to derive
the exact solution of the linear density evolution with all relativistic
corrections included,  valid at both the
sub-horizon and super-horizon, for 
 arbitrary $\Lambda$ and $K$. 

For LSS study,
the most widely used gauge is the Newtonian gauge (or the conformal Newtonian
gauge). Hence we will perform the derivation of the exact solution in the
Newtonian gauge, and then discuss its implications on the LSS-based
cosmology. Some of the 
perturbation variables in the Newtonian gauge, such as the matter overdensity
$\delta_m$, are not gauge invariant. However, such
gauge freedom is not a problem, as long as we properly connect the
perturbation variables  in the Newtonian gauge to observables (refer to
\cite{Yoo09} for a comprehensive treatment). Furthermore, once the solution in
the Newtonian gauge is derived, it can be transformed to other gauges by
performing a coordinate transformation (e.g. \cite{Kodama84,Ma95}). For these
reasons, we  will stick to the Newtonian gauge throughout the paper.

After we derived the solution and were writing this manuscript, we found an
earlier work by Jai-Chan Hwang \cite{Hwang94}, who has derived equivalent
solutions in not only the Newtonian gauge, but also in other five gauges. Nevertheless, we feel the need to publish
our result, for a number of reasons. First, the solutions are derived in independent
ways. \cite{Hwang94} first derived the solution in the comoving gauge and then
converted to  the Newtonian gauge (denoted as the zero
shear gauge in \cite{Hwang94}).  For us, we derive the solution directly in
the Newtonian 
gauge. Our result thus serves as an independent check and the
agreement between the two  verifies the validity of both solutions.  Second,
given recent discussions on the role of these relativistic
corrections (e.g. \cite{Dent09}),  it is useful to address the existence of these exact solutions
and clarify some confusions arisen in the literature. Third, the exact solution
derived has  direct applications to modern cosmological topics, some of them,
e.g. the primordial non-Gaussianity,  will be
elaborated later in the paper.  

The exact analytical solution that we found is valid for any
$\Lambda$ and curvature $K$. But numerical results and related
discussions  are  based on  the standard flat $\Lambda$CDM with
$\Omega_0=0.268$,  
$\Omega_\Lambda=1-\Omega_0$ and $\Omega_K=0$ throughout the paper, unless
otherwise specified.

\section{The exact solution}
We work with the FRW background metric 
\ba
ds^2&\equiv&-g_{\mu\nu}dx^{\mu}dx^{\nu}=
-dt^2+a^2\gamma_{ij}dx^idx^j \\
&=&-dt^2+a^2\left(\frac{dr^2}{1-Kr^2}+r^2d\theta^2+r^2\sin^2\theta
d\varphi^2\right) \nonumber\ .
\ea
The Hubble expansion rate is given by 
\be
\label{eqn:H}
H^2=\frac{8\pi G}{3}\left(\bar{\rho}_m+\rho_\Lambda\right)-\frac{K}{a^2}\ .
\ee
Here $\bar{\rho}_m$ is the mean cosmological matter density. 
Taking the derivative of both sides, we obtain a relation which will be used
later,
\be
\label{eqn:K}
K=4\pi G\bar{\rho}_ma^2+HH^{'}a^3\ . 
\ee
Throughout the paper, $^{'}\equiv d/da$. 

For reviews of linear perturbation in $\Lambda$CDM cosmology, please refer
to, e.g, \cite{Kodama84,Ma95,Bertschinger01}. Here we just summarize relevant
results.  For the convenience of LSS study, we adopt the Newtonian gauge
defined through 
\be
ds^2=-(1+2\phi)dt^2+a^2(1-2\phi)\gamma_{ij}dx^idx^j\ .
\ee
 Here, we have adopted
the result that the two Newtonian potentials are identical in $\Lambda$CDM
(e.g. \cite{Ma95}).  We choose the following three independent perturbation
equations,  
\be
\label{eqn:Poisson1}
(\nabla^2+3K)\phi-3a^2H^2(\phi^{'}a+\phi)=4\pi G\bar{\rho}_ma^2\delta_m\ ,
\ee
\be
\label{eqn:theta}
aH(\phi^{'}a+\phi)=4\pi G\bar{\rho}_m a^2W\ ,
\ee
\be
\label{eqn:delta}
\delta^{'}_m=\frac{\nabla^2W}{a^2H}+3\phi^{'}\ .
\ee
Here, $\nabla^2\equiv \gamma^{ij}\nabla_{;i}\nabla_{;j}$ and the covariant
derivative $\nabla_{;i}$ is defined in the 3D constant curvature space. $\vec{v}=-\nabla W$
is the peculiar velocity and $W$ is the velocity potential.  The three
equations above compose a complete set for the evolution of the three perturbation
variables, the Newtonian potential $\phi$, the matter overdensity $\delta_m$
and the velocity potential $W$.   

The $\phi^{'}a+\phi$ term in
Eq. \ref{eqn:Poisson1} is a relativistic correction to the Poisson equation. 
The $\phi^{'}$ term in Eq. \ref{eqn:delta} reflect the ambiguity to
distinguish inhomogeneities in space-time metric to that in the matter-energy
fluid. These relativistic corrections are usually neglected in LSS study. It
is the goal of this paper to quantify their impacts. 

\subsection{Linear evolution of the gravitational potential and peculiar velocity}
Interestingly, including all these relativistic terms allows us to
derive the exact solution. From Eq. \ref{eqn:Poisson1}, we obtain $\delta^{'}_m$,
\be
\delta^{'}_m=\left[\frac{(\nabla^2+3K)\phi-3a^2H^2(\phi^{'}a+\phi)}{4\pi
    G\bar{\rho}_ma^2}\right]^{'}\ .
\ee
Plug the above equation and Eq. \ref{eqn:theta} into Eq. \ref{eqn:delta}, we
obtain
\be
\left[\frac{(\nabla^2+3K)\phi-3a^2H^2(\phi^{'}a+\phi)}{4\pi G\bar{\rho}_ma^2}\right]^{'}=\frac{\nabla^2(\phi^{'}a+\phi)}{4\pi
  G\bar{\rho}_m a^3}+3\phi^{'}
\ee
Since $\bar{\rho}_m\propto a^{-3}$, in the above equation, the two terms
$\propto \nabla^2$ cancel exactly. We then have
\be
\left[\frac{3K\phi-3a^2H^2(\phi^{'}a+\phi)}{4\pi
    G\bar{\rho}_ma^2}\right]^{'}=3\phi^{'}
\ee
The solution is
\be
\label{eqn:phiK}
\frac{a^2H^2(\phi^{'}a+\phi)-K\phi}{4\pi G\bar{\rho}_ma^2}+\phi={\rm constant}\ .
\ee
Immediately we see the linear evolution in $\phi$ is scale independent. This
tells us that the solution of $\phi$ obtained in the sub-horizon limit is also
valid at super-horizon scale.

Multiple both sides by $4\pi G\bar{\rho}_ma^3/H^3a^3$ and use Eq. \ref{eqn:K},
we obtain 
\be
\left(\frac{a\phi}{H}\right)^{'}\propto \frac{1}{H^3a^3}\ .
\ee
Here, we have used the fact that $4\pi G\bar{\rho}_ma^3$ is a constant. 

The above equation has two independent solutions. The
solution $\phi\propto H/a$ corresponds to the decay mode in density. So it is
of little practical interest and will not be discussed hereafter. The other
solution corresponds to the growth mode in the density evolution, 
\be
\label{eqn:Dphi}
\phi\propto D_{\phi}\propto \frac{H}{a}\int_0^a \frac{da}{H^3a^3}\ .
\ee
Now the scale independence in $D_\phi$ is explicitly shown. 

The linear evolution in velocity (divergence) can be obtained combining
Eq. \ref{eqn:theta}  \& \ref{eqn:Dphi}, 
\be
W\propto D_v\propto aH^2\left(\frac{H^{'}a}{H}+\frac{1/H^3a^2}{\int_0^a
    da/H^3a^3}\right)\int_0^a \frac{da}{H^3a^3}\ .
\ee
Here, 
\be
\frac{H^{'}a}{H}+\frac{1/H^3a^2}{\int_0^a
    da/H^3a^3}=\frac{d\ln \phi}{d\ln a}+1 \ .
\ee
For the standard $\Lambda$CDM,  this quantity approaches unity at high redshift
and decreases to $\sim 0.5$ at $z=0$.

We notice that both the evolution in the gravitational potential and in the
velocity are identical to the ones derived in the sub-horizon (Newtonian) limit, namely,
\be
\label{eqn:phiv}
D_{\phi}=D_{\phi,N}\ \ ;\ \ D_v=D_{v,N}
\ee
Indeed, this is what expected from the scale independence in $D_\phi$ and
$D_v$. 

\bfi{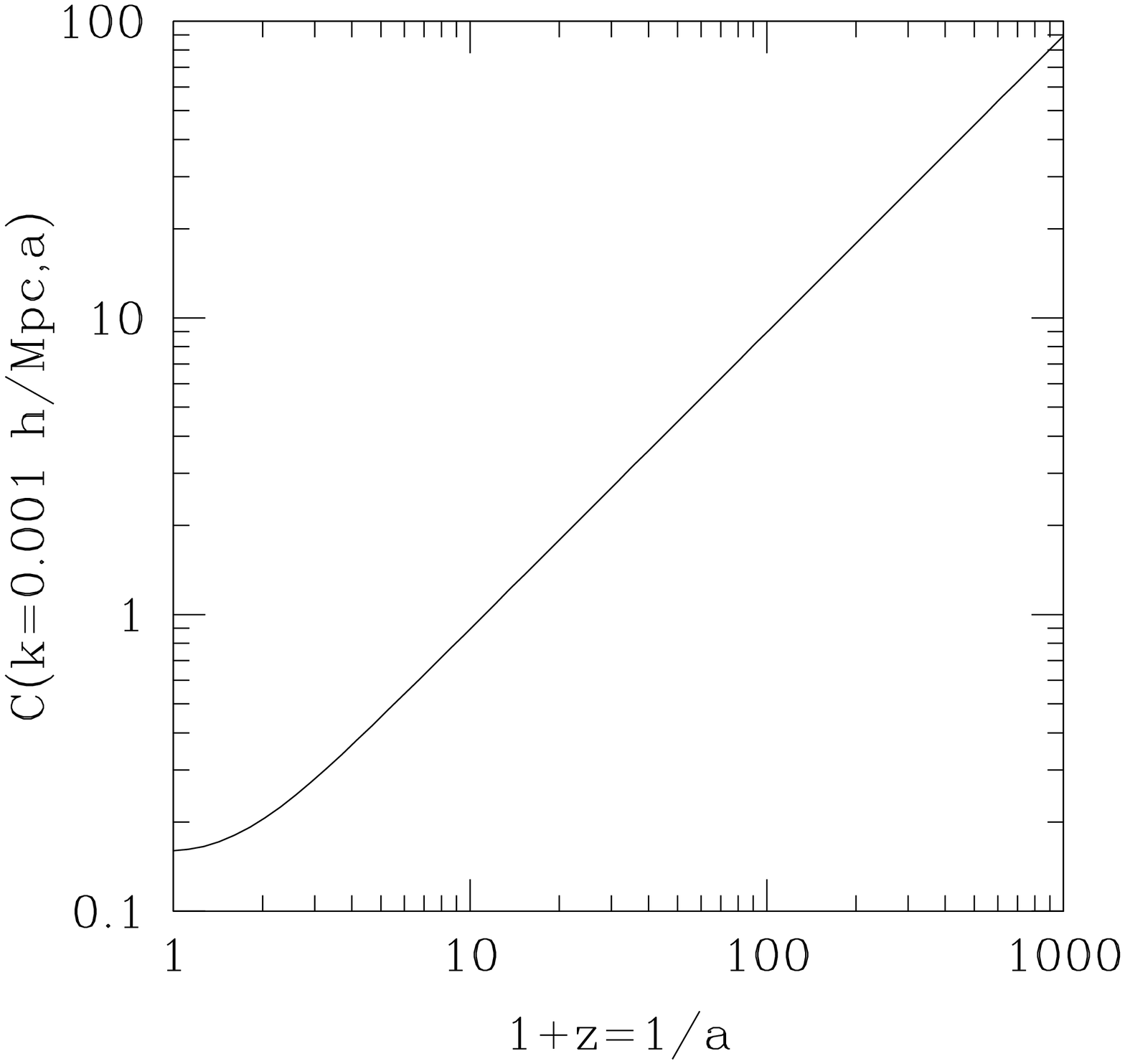}
\caption{The relativistic correction $C(k,a)$. Relativistic corrections alter
  the matter power spectrum by a factor $(1+C)^2$. So a positive $C$ means the
  amplification of the matter clustering. Since $C\propto k^{-2}$, we
  only plot $C(k,a)$ at $k=0.001 h/$Mpc. At $z\gg 1$, $C(k,a)\propto a^2$
  since $d\ln \phi/d\ln a\rightarrow 0$. \label{fig:C}}
\efi

\subsection{Linear evolution of the matter overdensity}
However, the situation for the overdensity evolution is different. From
Eq. \ref{eqn:Poisson1},  
\ba
\delta_m&=&\frac{2\Omega_0a}{3H_0^2}\times \\
&& \left[\nabla^2+3K-3a^2H^2\left(\frac{H^{'}a}{H}+\frac{1/H^3a^2}{\int_0^a
    da/H^3a^3}\right)\right]\phi\ . \nonumber
\ea
Only when the scale is much smaller than the curvature radius  and the horizon
scale, we recover Eq. \ref{eqn:Heath77}.

CMB observations, along with other probes,  show the universe to be nearly
flat ($|\Omega_K|\ll  1$) \cite{flat}.    Hence all modes accessible to
observations are much smaller than the curvature radius. We then
neglect the curvature term and replace $\nabla^2$ with the usual Laplace
operator in the 3D Euclid space. We further proceed to the Fourier space
($\nabla^2\rightarrow -k^2$). We denote the corresponding Fourier components
with a superscript ``$\sim$''.   We then have
\ba 
\label{eqn:tildedeltam}
\tilde{\delta}_m=-\frac{2\Omega_0a}{3H_0^2}k^2\tilde{\phi}\times \left[1+C(k,a)\right]\ . 
\ea
The impact of relativistic corrections is then completely captured by the
relativistic correction  term
\ba
C(k,a)&=&\frac{3a^2H^2}{k^2}\left(\frac{H^{'}a}{H}+\frac{1/H^3a^2}{\int_0^a
    da/H^3a^3} \right)\\
&=&\frac{a^2(H/H_0)^2}{3(k\times 10^3\mpch)^2}\left(\frac{H^{'}a}{H}+\frac{1/H^3a^2}{\int_0^a
    da/H^3a^3} \right)\nonumber\ .
\ea
For the standard $\Lambda$CDM cosmology, $C>0$. The numerical result is plotted in Fig. \ref{fig:C}.

\bfi{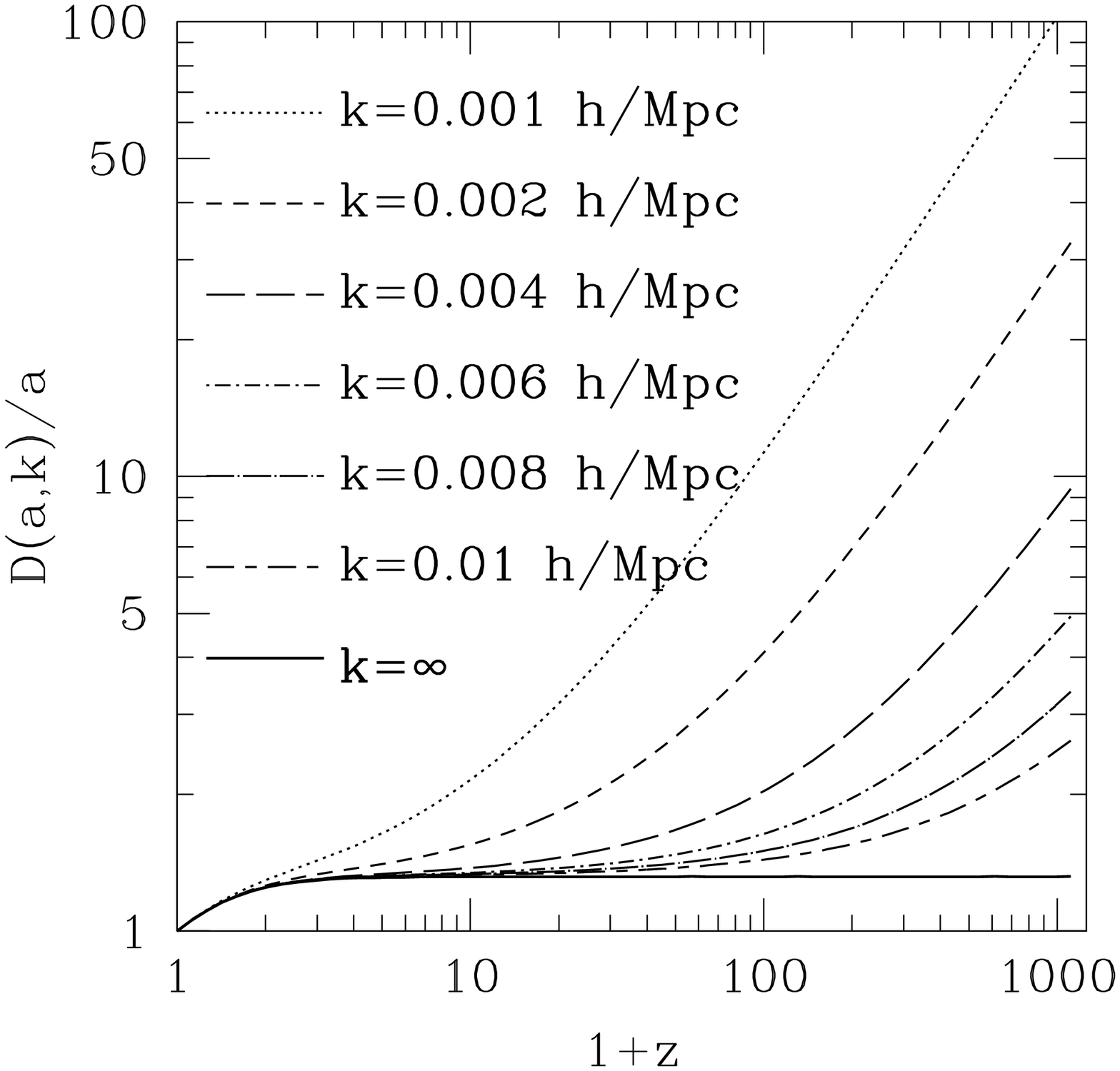}
\caption{The linear density evolution in $\Lambda$CDM under the Newtonian
  gauge, for various $k$. The solid line corresponds to the one in the
  Newtonian limit.  Larger deviation from the Newtonian result happen at
  higher redshifts, where the relativistic corrections are larger due to
  smaller horizon scale. \label{fig:D}}
\efi

The matter power spectrum $P_m$ is related to the power spectrum of the
gravitational potential $P_\phi$ through Eq. \ref{eqn:tildedeltam}, 
\ba
P_m(k,z)&=&\left[\frac{2\Omega_0a}{3H_0^2}\right]^2k^4P_\phi\times
\left[1+C(k,a)\right]^2 \\
&=& P_{m,N}(k,z)\times \left[1+C(k,a)\right]^2\ . \nonumber
\ea
 Since $C>0$, relativistic corrections enhance the matter clustering. 

The linear density evolution is derived combining Eq. \ref{eqn:tildedeltam} \&
\ref{eqn:Dphi}, 
\ba
\tilde{\delta}_m\propto \left[H\int_0^a \frac{da}{H^3a^3}\right] \times \left[1+C(k,a)\right]\ .
\ea

Relativistic corrections induce a characteristic scale
dependence $k^{-2}$ in the otherwise scale independent density
growth in $\Lambda$CDM cosmology. Only when $k\gg aH$, does this 
scale dependence vanishes ($C(k,a)\ll 1$).
The linear density growth factor $D_m$ is often defined as $D_m\equiv
\delta_m(a)/\delta_m(a=1)$ and normalized at $z=0$, in LSS study. We then have
\ba
\label{eqn:D}
D_m(a,k)&=&\left[\frac{H(a)}{H_0}\right]\left[\frac{\int_0^a da/H^3a^3}{\int_0^1
    da/H^3a^3}\right]\left[\frac{1+C(k,a)}{1+C(k,a=1)}\right] \nonumber \\
&=& D_{m,N}(a)\times \left[\frac{1+C(k,a)}{1+C(k,a=1)}\right] \ .
\ea

\bfi{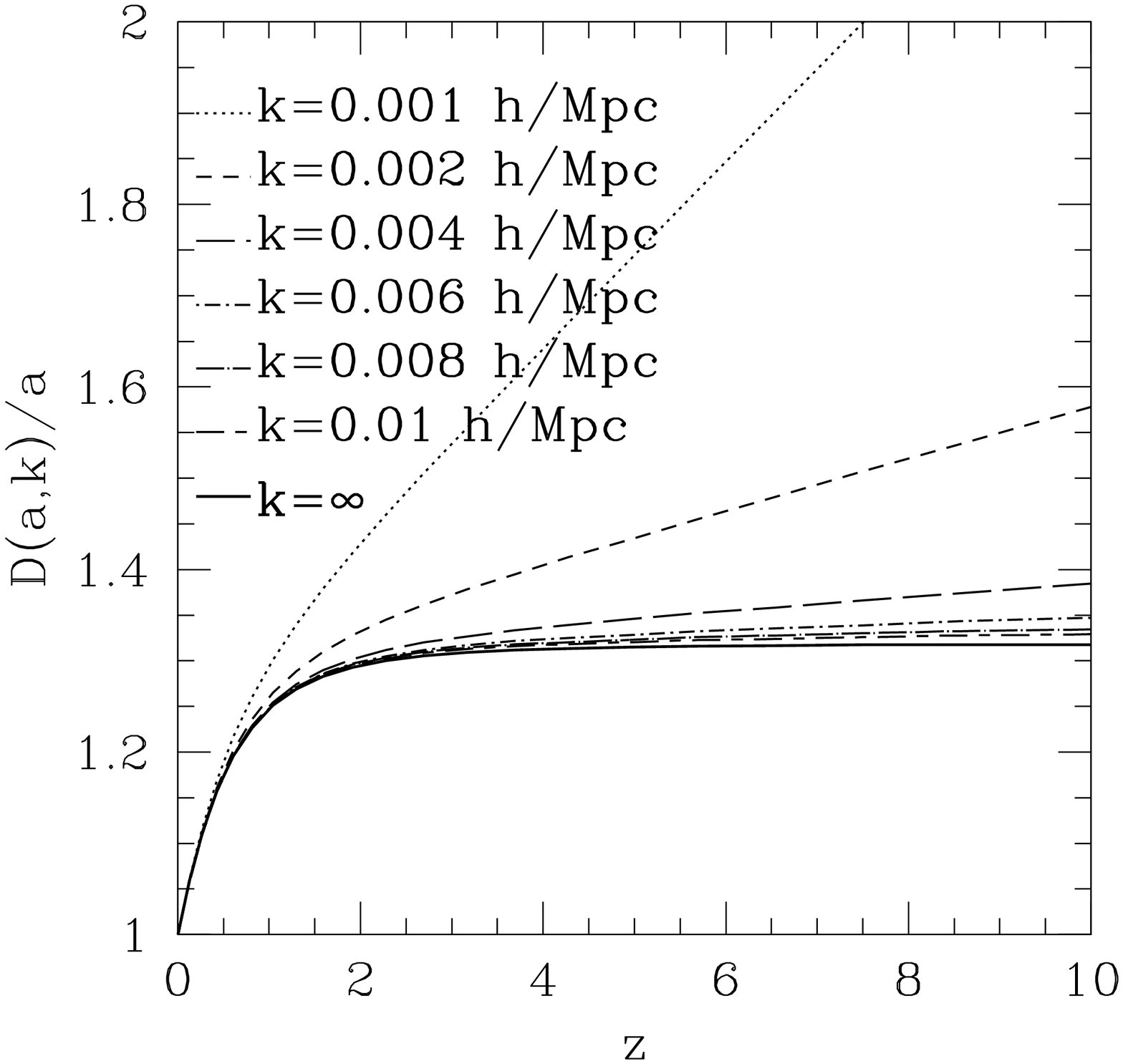}
\caption{Same at Fig. \ref{fig:D}, but only show the evolution at $z<10$ where
  LSS surveys may have access. \label{fig:D1}}
\efi
We plot $D(a,k)/a$ for various $k$ in Fig. \ref{fig:D} \& \ref{fig:D1} . As expected, the
relativistic corrections become bigger at higher redshifts and can completely
invalidate the Newtonian result (Eq. \ref{eqn:Heath77}), due to shrinking
horizon. This analytical result verifies the finding of \cite{Dent09},  who
used numerical calculation to quantify the relativistic corrections.  One thing to notice is that, \cite{Dent09}  normalizes the
overdensity at $z=1100$ and hence defines $D=\delta_m(z)/\delta_m(z=1100)$, so
deviations from the Newtonian limit appear at lower redshift instead. 

Surprisingly, the large deviation shown in Eq. \ref{eqn:Heath77} and Fig. \ref{fig:D}
does not necessarily invalidate the applicability of the Newtonian approximation
(Eq. \ref{eqn:Heath77}) in LSS study. This is essentially a normalization
issue. To better demonstrate this point, we highlight the linear density
growth at $z<10$ in Fig. \ref{fig:D1}. Most LSS surveys are limited to $z\la
4$.  If we solely compare LSS  
at redshifts accessible to those surveys to measure the structure growth rate,
Eq. \ref{eqn:Heath77} is essentially exact for $k>0.01h/$Mpc. Even for the mode $k=0.006h/$Mpc, the accuracy is better than $1\%$, negligible
comparing to  the cosmic variance.  Hence in general, Eq. \ref{eqn:Heath77}
is sufficiently 
accurate to describe the linear growth at redshifts accessible to LSS
surveys at $z\la 4$. However, later in the paper we will discuss an important
exception, where by the multiple tracer technique proposed by \cite{Seljak09},
the cosmic variance can be eliminated.

21 cm surveys can probe LSS to the reionization epoch at  $z\sim 10$ and even
higher \cite{21cm}. A comparison between these epochs and $z\sim 0$ based on
the Newtonian approximation (Eq. \ref{eqn:Heath77}) would lead to $\ga 1\%$ error
in the derived density growth rate at $k=0.01 h/$Mpc, still small, but may no
longer be negligible. 

On the other hand, if we want to compare LSS measurements at low redshift to
CMB at 
$z\simeq 1100$, Eq. \ref{eqn:Heath77} is no longer applicable, shown in Fig. \ref{fig:D}. Even for
$k=0.01h/$Mpc, the error induced is a factor of $2$.  Nevertheless, there are
other sources of error prohibiting the application of Eq. \ref{eqn:Heath77} in
such case, such as non-negligible radiation and non-identical baryon and dark
matter distribution at $z\ga 100$. CMB packages such as CMBFAST and CAMB have already taken these
complexities into account and shall be used to compare between high and low
redshift observations. 

\bfi{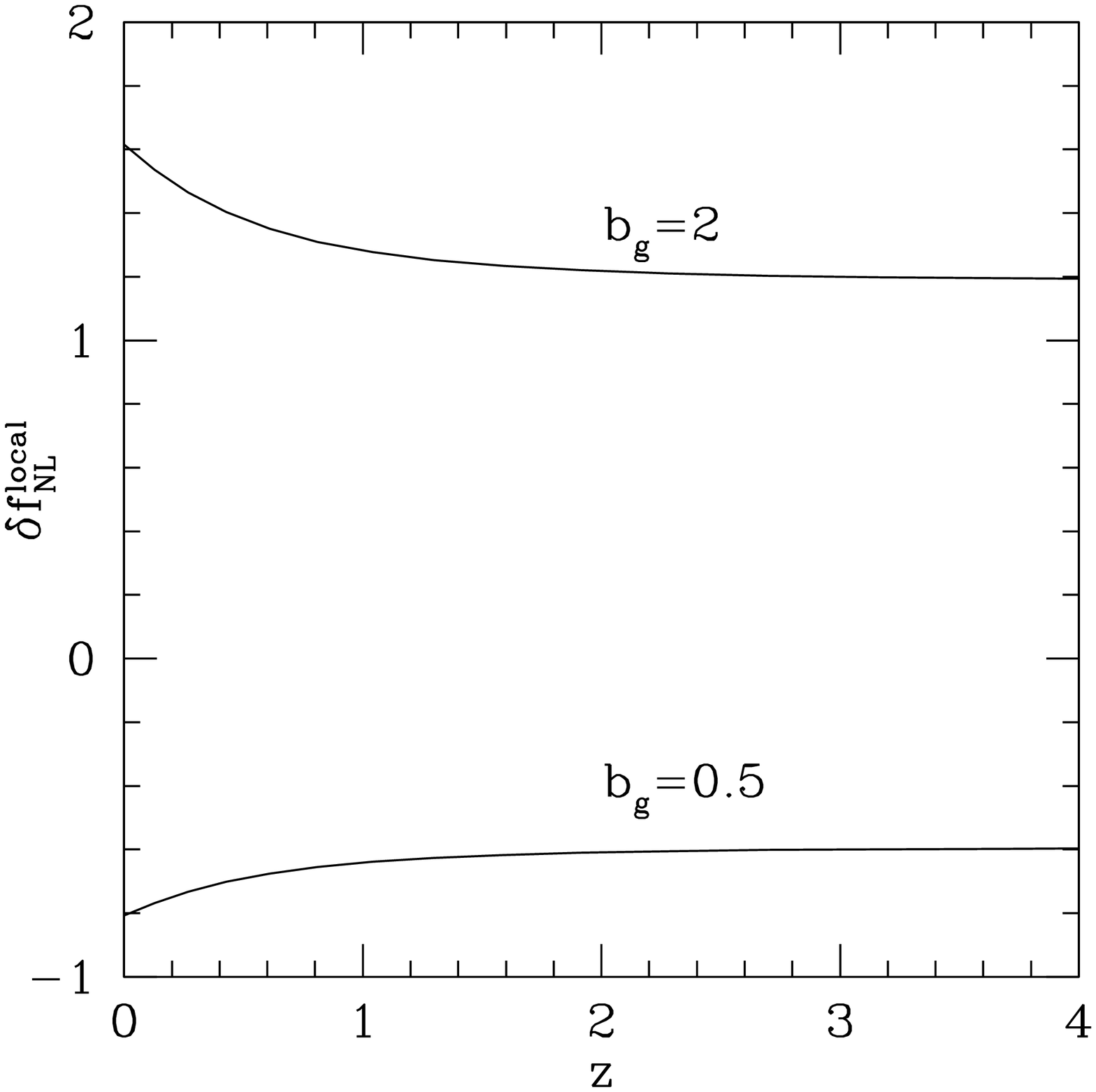}
\caption{General relativistic corrections to the Poisson equation and the
  continuity equation can mimic a primordial non-Gaussianity of local type and
  induce a bias $\delta f_{\rm NL}^{\rm local}\propto b_g/(b_g-1)$. The plot
  shows the case for $b_g=2$ and $b_g=0.5$ respectively.  \label{fig:fnl}}
\efi

\section{Cosmological impacts of relativistic corrections}
In this section we discuss the impact of these relativistic corrections on a
number of cosmological applications of LSS.  We find that non-negligible
impact may occur for searching for primordial non-Gaussianity through
two-point galaxy clustering and N-body simulations of Gpc box size. 

\subsection{Primordial non-Gaussianity}
The  galaxy overdensity $\delta_g$ is an observable and hence gauge independent. In the
Newtonian gauge, it can be expressed as  \cite{Yoo09}
\be
\label{eqn:deltag}
\delta_g=b_g\delta_m+f(\phi,\vec{v})\ .
\ee
Here, $f$ is a linear function of $\phi$ and $\vec{v}$. We have shown that the
relativistic corrections discussed earlier do not bias the Newtonian
calculation on $\phi$ and  $\vec{v}$, so the only term affected  by the
relativistic corrections is the usual
$b_g\delta_m$ term, where $b_g$ is the galaxy bias. 

Primordial non-Gaussianity (PMG) of the local type  induces 
scale dependence to the otherwise scale-independent bias at large scale
\cite{Dalal08}, 
\be
\label{eqn:Deltabg}
\Delta b_g=2(b_g-1)f^{\rm local}_{\rm NL}
\delta_c\frac{3\Omega_mH_0^2}{2ag(a)k^2}\ .
\ee
Here, $f^{\rm local}_{\rm NL}$ is the PMG parameter. $\delta_c$ is the threshold for halo collapse and $g\propto
D(k\rightarrow \infty)/a$, normalized such that $g(a\rightarrow 0)\rightarrow
1$. If we miss the relativistic 
corrections in the matter evolution,  it will bias the interpretation of
galaxy bias by 
\be
\delta b_g=b_gC(k,a)\ .
\ee
This in turn causes a systematical error in $f^{\rm local}_{\rm NL}$,
\ba
\delta f^{\rm local}_{\rm NL}&=&\frac{b_ga^3H^2g(a)}{(b_g-1)\delta_c\Omega_0H_0^2}\left(\frac{H^{'}a}{H}+\frac{1/H^3a^2}{\int_0^a
    da/H^3a^3} \right)\nonumber \\
&=& \frac{b_g}{b_g-1}\times O(1) \ \ .
\ea
It diverges when $b_g\rightarrow 1$. But this divergence is
trivial, simply meaning  that 
PMG of the local type does not affect the two-point clustering of  $b_g=1$ galaxies
(Eq. \ref{eqn:Deltabg} and more detailed discussions in \cite{Dalal08}). The induced biases for $b_g=2$ and $b_g=0.5$ are shown in Fig. \ref{fig:fnl}. 
Existing constraints on $f^{\rm local}_{\rm NL}$ from two-point galaxy clustering \cite{Slosar08} are
at least one order of magnitude larger, so the bias induced by relativistic
corrections is negligible for existing surveys and stage IV surveys such as LAMOST
\cite{Gong09} and BOSS.   However, deep full sky LSS surveys have the potential to
constrain  $f^{\rm local}_{\rm NL}$ with statistical error $\Delta f^{\rm
  local}_{\rm NL}=O(1)$
\cite{Seljak09}. For these surveys, 
the relativistic correction induced bias should  be included in the analysis. 

Terms linear in $\phi$ in Eq. \ref{eqn:deltag} also induce bias to $f^{\rm local}_{\rm NL}$ \cite{Yoo09,Yoo10}.
The two systematical biases in $f^{\rm local}_{\rm NL}$ are independent and comparable in
amplitude.

\subsection{Large box N-body simulation}
Large simulation box size $L\sim 1\gpch$ is required to calculate the
nonlinear matter power spectrum at $1\%$ level accuracy
\cite{Heitmann10}. This is also the requirement for primordial
non-Gaussianity study.  N-body simulation usually adopts the Poisson equation
$\nabla^2 \phi=4\pi G\bar{\rho}_ma^2\delta_m$ to calculate the potential and
hence misses relativistic corrections. The
largest modes accessible to these simulations are $k\sim 2\pi/L\sim
0.006h/$Mpc.  N-body simulations typically begin at $z=100$. For the largest
modes, the density evolution  to $z=0$ is then biased by more than $20\%$,
definitely non-negligible. 

This problem can be solved in the
post-processing of simulation. Eq. \ref{eqn:phiv} means that the simulated
gravitational potential $\phi$ is correct. So from Eq. \ref{eqn:Poisson1}, by simply
adding  a term $-2a^3(H/H_0)^2(\phi^{'}a+\phi)/\Omega_0$ to the simulated
$\delta_m$, we can recover the correct density evolution. 

\subsection{Redshift distortion} 
Redshift distortion is emerging as a promising probe of the large scale
structure \cite{RD}. Cosmological information encoded in redshift distortion is often
quantified by a single parameter $f$.  The usual definition is $f\equiv d\ln
D_m/d\ln a$.  Through this definition, one can work out the impact of
relativistic corrections straightforwardly.

Alternatively, one can define it as 
\ba
f&\equiv & \frac{-\theta/aH}{\delta_m}\equiv \frac{\nabla^2W/aH}{\delta_m}\ .
\ea
This definition does not assume a deterministic peculiar velocity-overdensity
relation and is hence more general. We then have
\ba 
f=\frac{H^{'}a/H+1/H^3a^2\int_0^a
    da/H^3a^3}{1+C(k,a)}=\frac{f_N}{1+C(k,a)} \nonumber\ .
\ea
Here, $f_N$ is the $f$ in the Newtonian limit. Relativistic corrections
induce scale dependence in $f$, quantified by $C(k,a)$.

The $\gamma$-index is often adopted to describe the structure growth \cite{gammaindex}.  If we
straightforwardly extend the usual definition, we have
\ba
\gamma(a,k)&=&\frac{\ln f}{\ln \Omega_m(a)} \\ 
&=&\gamma_N(a)-\frac{\ln [1+C(k,a)]}{\ln
  \Omega_m(a)}\simeq \gamma_N-\frac{C(k,a)}{\ln
  \Omega_m(a)}\nonumber\ .
\ea

For  most scales and redshifts accessible to the precision redshift distortion
measurement,  $C\sim (k\times 10^3\mpch)^{-2}\ll 1$.  So relativistic
corrections have negligible impact on $f$ and $\gamma$. 

\section{Summary}
We consider relativistic corrections to the Poisson equation
and the continuity equation and investigate their impact to the linear
evolution of perturbations in $\Lambda$CDM. Major results are as follows. 
\bi
\item We derive  the exact analytical solution for the
linear evolution of gravitational potential, velocity and matter overdensity under
the Newtonian gauge.  It is valid at both the sub- and super-horizon scale,
for arbitrary cosmological constant and curvature. 
\item Relativistic corrections amplify the matter clustering and  introduce a $k^{-2}$ scale dependence to the
  density fluctuation.  The overdensity evolution is altered by $10\%$ at
  $k=0.01h/$Mpc from $z=100$ to $z=0$. 
\item The gravitational potential and velocity evolution obtained under the
Newtonian limit remaines exact with the presence of these relativistic
corrections. 
\item  If these relativistic  corrections are not included in analyzing the
  two-point galaxy clustering,  constraint on the primordial non-Gaussianity
  will be biased by $\delta f^{\rm  local}_{\rm NL}\sim 1$. 
\item These relativistic corrections bias N-body simulations, which are based on the
  Newtonian approximation. For $\sim \gpch $ box size simulations, the largest
  mode of density fluctuations is underestimated by $\sim 20\%$ from $z=100$
  to $z=0$. We further show that this effect is correctable in the simulation
  post-processing. 

\ei

One question remains is whether we can extend the analytical solution to dark
energy models. Under the limit that dark energy fluctuation is negligible,
Eq. \ref{eqn:Poisson1}, \ref{eqn:theta} \& \ref{eqn:delta} hold. So it seems
that we can repeat the derivation for smooth dark energy models up to
Eq. \ref{eqn:phiK}, which allows us to express the linear evolution with a double
integral form.  Unfortunately, numerically we find that the solution obtained
in this way does not reproduce the known (and correct) behavior at deep
sub-horizon scale.  The reason is that, dark energy fluctuations are
inevitable since $w\neq -1$ \footnote{Gravitational potential sources dark
  energy fluctuation through a term $(1+w)\dot{\phi}$ \cite{Ma95}. Hence unless $w=-1$,
  dark energy fluctuation is inevitable. More rigorous proof can be achieved
  by resorting to the full linearized energy-momentum conservation equations in
  \cite{Ma95}. }.  It is true that these fluctuations can be small
and negligible  with respect to the dominant terms in Eq. \ref{eqn:Poisson1},
\ref{eqn:theta} \& \ref{eqn:delta}. However, they are comparable to $\phi$ in
Eq. \ref{eqn:phiK}.  By neglecting them at first hand in Eq. \ref{eqn:Poisson1},
\ref{eqn:theta} \& \ref{eqn:delta}, we miss non-negligible corrections
from dark energy fluctuations to Eq. \ref{eqn:phiK}.

\section{Acknowledgment}
I thank Jun Zhang and Xuelei Chen for many helpful discussions. This work is
supported by the one-hundred talents program of the Chinese 
academy of science, the national science 
foundation of China (grant No. 10821302, 10973027\& 11025316),   the CAS/SAFEA
International Partnership Program for  Creative Research Teams and National
Basic Research Program of China (973 Program) under grant No.2009CB24901.


\begin{thebibliography}{}
\bibitem{Heath77} D.J. Heath, MNRAS, 179, 351 (1977)
\bibitem{Dalal08}  Neal Dalal, Olivier Doré, Dragan Huterer, Alexander Shirokov. 	Phys.Rev.D77:123514
  (2008). [arXiv:0710.4560] 
\bibitem{Dent09} James B. Dent, Sourish Dutta. Phys. Rev. D 79, 063516
  (2009). [arXiv:0808.2689]

\bibitem{CMBFAST} Uros Seljak, Matias Zaldarriaga. ApJ, 469, 437 (1996) 
\bibitem{CAMB} A. Lewis, A. Challinor. http://camb.info/
\bibitem{Yoo09} Jaiyul Yoo, A. Liam Fitzpatrick, Matias Zaldarriaga.
  Phys.Rev.D80:083514 (2009). [arXiv:0907.0707]


\bibitem{Ma95} Chung-Pei Ma, Edmund Bertschinger. Astrophys.J. 455 (1995) 7-25 
\bibitem{Kodama84} Kodama, Hideo; Sasaki, Misao. Progress of Theoretical
  Physics Supplement, Vol. 78, p. 1, 1984 

\bibitem{Hwang94} Hwang, J.-C.\ 1994, \apj, 427, 533 
\bibitem{Bertschinger01} Edmund Bertschinger. arXiv:astro-ph/0101009

\bibitem{flat} P. de Bernardis, et al. Nature, 404, 955 (2000); E. Komatsu, et
  al.  arXiv:1001.4538 (2010)

\bibitem{Seljak09} Uros Seljak. Phys.Rev.Lett.102:021302 (2009)
\bibitem{21cm} e.g. Abraham Loeb, Matias Zaldarriaga. PRL, 92, 211301 (2004);
  U. Pen, New Astron. 9, 417 (2004); 
  Xuelei Chen, Jordi Miralda-Escude. ApJ, 684,18 (2008) [arXiv:astro-ph/0605439]
  Pengjie Zhang, Zheng Zheng, Renyue Cen. MNRAS, 382, 1087 (2007) 



\bibitem{Slosar08} Anze Slosar, Christopher Hirata, Uros Seljak, Shirley Ho,
  Nikhil Padmanabhan. 	JCAP 08, 031 (2008)
\bibitem{Gong09} Yan Gong, Xin Wang, Zheng Zheng, Xuelei
  Chen. Res.Astron.Astrophys.10, 107 (2010) 

\bibitem{Yoo10} Jaiyul Yoo. arXiv:1009.3021 (2010)
\bibitem{Heitmann10} Katrin Heitmann, Martin White, Christian Wagner, Salman
  Habib, David Higdon. Astrophys.J.715:104-121 (2010). [arXiv:0812.1052]
\bibitem{RD}  e.g. U.Pen. ApJ, 504, 601 (1998); J. Peacock, et al. Nature 410, 169-173 (2001); Max Tegmark,
  Andrew J. S. Hamilton, Yongzhong
  Xu. MNRAS, 335, 887 (2002); M. Tegmark, et al. ApJ, 606, 702 (2004); Pengjie Zhang, Michele Liguori, Rachel Bean, Scott
  Dodelson. PRL, 99, 141302 (2007);  Viviana Acquaviva, Amir Hajian, David
  N. Spergel, Sudeep Das. PRD, 78, 043514 (2008); Bhuvnesh Jain, Pengjie
  Zhang. PRD, 78, 063503 (2008);   L. Guzzo, et al. Nature 451:541-545,2008; Pengjie Zhang, Rachel Bean, Michele
  Liguori, Scott Dodelson. arXiv:0809.283 (2008); Eric
  V. Linder. Astropart.Phys.29:336 (2008);  Yun Wang. JCAP, 0805,021(2008);
  Martin White, Yong-Seon Song, Will 
  J. Percival. MNRAS, 397,1348 (2008); Reinabelle Reyes, Rachel
  Mandelbaum, Uros Seljak, Tobias Baldauf, James E. Gunn, Lucas Lombriser,
  Robert E. Smith. Nature, 464, 256-258 (2010)

\bibitem{gammaindex} P.J.E. Peebles. ApJ, 205, 318 (1976); Limin Wang, Paul
  J. Steinhardt. ApJ, 508, 483 (1998); E.V. Linder. PRD, 72, 043529 (2005)
\end{thebibliography}
\end{document}